\documentclass[10pt,A4letter]{article}             % onecolumn (standard format)
\usepackage{natbib}

\usepackage{fourier}
\usepackage{graphicx}
\usepackage{amsmath}
\usepackage{amsfonts}
\usepackage[latin1]{inputenc}
\usepackage[T1]{fontenc}
\usepackage{verbatim}
\usepackage{latexsym}

\usepackage[backref]{hyperref}
\hypersetup{colorlinks=true,citecolor=cyan,urlcolor=cyan,linkcolor=blue}

\begin{document}

\sloppy

\title{Methods for characterising microphysical processes in plasmas}

\author{T. Dudok de Wit$^1$, O. Alexandrova$^2$, I. Furno$^3$, L. Sorriso-Valvo$^4$, G. Zimbardo$^5$}

\date{\footnotesize
$^1 $ LPC2E, CNRS and University of Orl\'eans, \\
3A avenue de la Recherche Scientifique, 45071 Orl\'eans cedex 2, France\\
$^2 $  O. Alexandrova, LESIA, Observatoire de Paris \\
5, place Jules Janssen, 92190 Meudon, France\\
$^3 $  I. Furno, EPFL SB CRPP,  Station 13, 1015 Lausanne, Switzerland\\
$^4 $  L. Sorriso-Valvo, Dipartimento di Fisica,IPCF-CNR, \\ UOS di Cosenza, 
Ponte Pietro Bucci Cubo 31C,
87036 - Arcavacata di Rende (CS), Italy\\
$^5 $   G. Zimbardo, Dipartimento di Fisica, Universit\`a della Calabria, \\
Ponte Pietro Bucci Cubo 31C, 87036 - Arcavacata di Rende (CS), Italy\\
}
\maketitle

\begin{abstract}
\textit{This is a slightly expanded version of an article to appear in Space Science Reviews (2013), doi: 10.1007/s11214-013-9974-9.} \\

Advanced spectral and statistical data analysis techniques have greatly contributed to shaping our understanding of microphysical  processes in plasmas. We review some of the main techniques that allow for characterising fluctuation phenomena in geospace and in laboratory plasma observations. Special emphasis is given to the commonalities between different disciplines, which have witnessed the development of similar tools, often with differing terminologies. The review is phrased in terms of few important concepts: self-similarity, deviation from self-similarity (\textit{i.e.} intermittency and coherent structures), wave-turbulence, and anomalous transport.
\end{abstract}

\section{Introduction}
\label{sec:intro}
Space and laboratory plasmas are fundamentally governed by both couplings across different scales and by nonlinear processes. This applies to a wide range of scales: from the fluid regime, where the magnetohydrodynamic (MHD) approximation can be used \citep{biskamp03}, down to small scales, where kinetic effects should be taken into account \citep{akhiezer75}. Progress toward their physical understanding has been closely associated with our ability to infer pertinent information by means of appropriate data analysis techniques. Usually,  physical space is very sparsely sampled in the sense that one has access only to few observables (magnetic field, electron density, plasma velocity and temperature, \ldots), often with poor temporal and/or spatial coverage. The challenge then consists in recovering the information of interest from highly scattered observations.

For many decades, the analysis of plasma data has been dominated by techniques that implicitly assume linearity or at best rely on second order moments only (\textit{e.g.} correlation functions and power spectra). Our prime objective here is to show how more advanced techniques can often provide deeper physical insight by extracting quantitative information that might otherwise have gone unnoticed. For example, two similarly-looking turbulent wavefields with identical power spectral densities may actually exhibit quite different properties in their higher order spectra, which in turn has direct implications on the existence of nonlinear wave couplings (or structures) that transfer energy between scales. 

Many new techniques have been developed in the two last decades, thanks to numerous advances in neighbouring fields such as dynamical systems \citep{bohr05,kantz00} or complex systems \citep{badii99}. Interestingly, most of these techniques are remarkably universal, and are used almost equivalently in laboratory and in astrophysical plasmas, and also in neutral fluids. Here, we discuss some of these concepts in the context of turbulence and large-scale structures in plasmas. Special emphasis will be given to commonalities between different disciplines, which are often concealed by differing nomenclatures. The focus will be on time series from geospace and from laboratory plasmas.

Regrettably, the literature is almost devoid of reviews that cover more than one discipline only. For laboratory plasmas, some relevant references are the works by \citet{ritz88,skoric08,tynan09,fujisawa10}. In geospace plasmas, besides a few general reviews \citep{wernik96,ddw03,bruno05,vassiliadis06,zimbardo10} a large effort has been directed towards multipoint data analysis \citep{paschmann00,paschmann08}.

This overview, rather than being exhaustive, will be phrased in terms of three frequently encountered concepts, namely 
self-similarity, intermittency and coherent structures (Sec.~\ref{sec:selfsimilar}), measurements of wave-turbulence in space and time (Sec.~\ref{sec:spatiotemporal}) 
and anomalous transport (Sec.~\ref{sec:anomalous}). Unfortunately we had to leave out several relevant and timely topics such as techniques for dynamical systems, or for image processing.

%%%%%%%%%%%%%%%%%%%%%%%%%%%%%%%%%%%%%%%%%%%%%%%%%%%%%%%%%%%%%%%%%%%%%%%%%%%%%%%%%%%%%%%%%%%%%%%%%%%%%%%%%%%%%%%%%%

\section{Self-similarity in turbulent plasmas}
\label{sec:selfsimilar}

Turbulence is ubiquitous in geospace and in laboratory plasmas, and occurs whenever energy injection occurs in the system and the dissipation mechanisms are weak (\textit{i.e.} efficient at much smaller scales than the scale of injection). Turbulent states are characterised by wavefield fluctuations covering all scales from the energy injection scale $L$ down to the scale $\ell_d$ at which energy dissipates. 

In neutral fluids, these fluctuations exhibit a power law spectrum $E(k) \sim k^{-5/3}$ in  Fourier space ($k$ being a wavenumber) between $k_L=1/L$ and $k_d=1/\ell_d$, called the {\it inertial range}.  Observation of this law for any turbulent flow indicates its universality and supports the idea of a self-similar behaviour of the turbulent velocity fluctuations, whose amplitude at scale $l$ varies as $\delta u_{l} \sim \delta u_L l^{\alpha}$. This is to be expected if the physical processes are not scale-dependent, namely if the dynamic equations (\textit{e.g.} Navier-Stokes for neutral fluids or MHD for magnetised plasmas) are invariant under scale transformation. This idea was used by Kolmogorov in 1941 to explain the ubiquitous $E(k) \sim k^{-5/3}$ spectrum \citep{kolmogorov1941}.  

Some time later, it was pointed out that the idea of self-similarity would imply spatial homogeneity of the process \citep{kolmogorov1962}. However, studies of higher order statistics show that this property is not verified for all turbulent flows. Departure from self-similarity is called intermittency in fully developed turbulence, and is another universal property of  most turbulent systems. Intermittency is attributed to the inhomogeneous energy flux through the scales, and results in the presence of small-scale coherent structures in the turbulent flow, in which most of the energy dissipation is concentrated. 
  
In this Section, we shall discuss first how to determine  energy spectra and related spectral indices from fluctuating wavefields in geospace and in fusion plasmas. Next we shall introduce techniques for analysing intermittency in time series.

%%%%%%%%%%%%%%%%%%%%%%%%%%%

\subsection{The solar wind as turbulence laboratory: power law determinations}\label{sec:spectra}

With its wide range of scales and parameters conditions, the solar wind is the best laboratory for studying plasma turbulence, thanks also to the availability of \textit{in situ} measurements that are provided by dedicated space missions \citep[\textit{e.g.},][]{bruno05}. Therefore, most of the following will refer to solar wind turbulence, unless stated otherwise.

Because of the large number of degrees of freedom in the dynamics of a turbulent flow, most of the results about turbulence refer to statistical quantities. As for any statistical study, in plasma turbulence it is important to check that ergodicity and stationarity apply. Ergodicity, \textit{i.e.} the equivalence between time and space averaging, is normally ensured if the sample size of the time interval $T$ is much larger than a typical correlation time, $T\gg t_c$.  The correlation time $t_c$ can be estimated from the data, for example as the time lag at which the autocorrelation function of the field vanishes to noise level, or reaches the value $1/e$. In turbulent flows, this should correspond to the integral scale, at which energy is injected in the system \citep{frisch95}. Ergodicity cannot be formally proven on the basis of observations, and yet, various studies suggest that it is a reasonable working hypothesis. For example, ensemble averages of the solar wind magnetic fields follow the asymptotic behaviour predicted by the ergodic theorem \citep{matthaeus82}.

Stationarity is usually supported by observations from laboratory plasmas when the discharge can be controlled, while for geospace plasmas it is often more difficult to extract stable and homogeneous records. Large scale, externally driven structures can indeed affect the statistical properties of the turbulent fields. For example, solar wind plasma properties change for fast and slow wind populations, each originating from specific regions of the solar atmosphere \citep{tu95,bruno05}.  Large-scale boundaries between these different regions may mix the phases and destroy the correlations that are present in nonlinear energy cascades. This is why the first important step towards the analysis of a turbulent signal is the choice of an appropriate time interval that excludes large scale boundaries, magnetic connection to planetary shocks, \textit{etc.} Meanwhile, many authors have studied long periods of solar wind data that included a mix of different regimes. Although this kind of analysis can give insight into the statistics of the field fluctuations, its interpretation in terms of phenomenological or theoretical modelling can be misleading, and should be carefully discussed.

\begin{figure}[!htb] 
\centering
\includegraphics[width=0.80\textwidth]{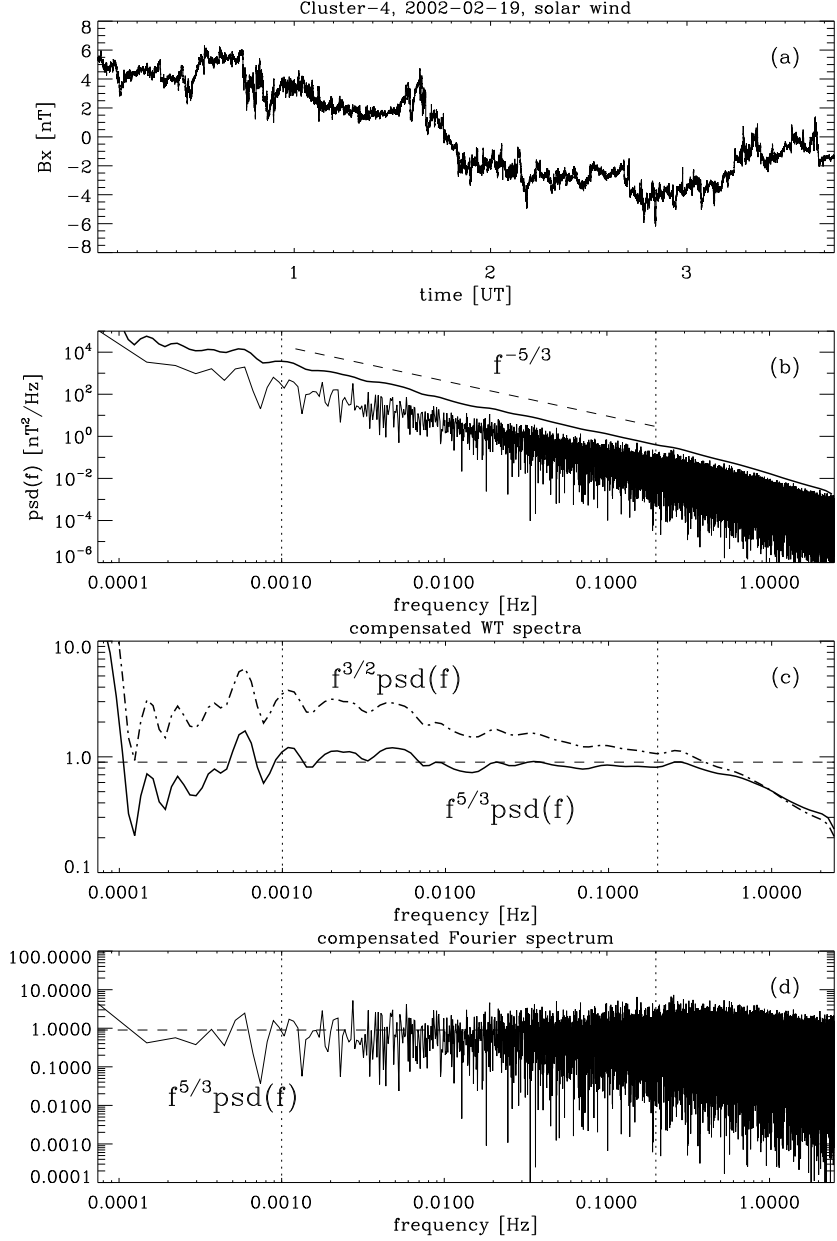}
\caption{Example of solar wind observations by CLUSTER. This time interval was used to study turbulence by \citet{bale05}. The plots show: (a) $B_x(t)$; (b) Fourier and Morlet wavelet spectra of $B_x(t)$, $f^{-5/3}$ is indicated by a dashed line, frequency range $\Delta f$ where the fitting was done is delimitated by two vertical dotted lines; (c) Compensated wavelet spectrum by $f^{5/3}$--function (solid line) and by $f^{3/2}$--function (dashed-dotted line); (d) Compensated Fourier spectrum by $f^{5/3}$--function. \label{fig:spec}
}
\end{figure}

Figure~\ref{fig:spec}(a) shows the temporal evolution of the $B_x(t)$ component of the magnetic field along the solar wind flow direction, as measured by the CLUSTER-4 spacecraft during the time interval studied by \citet{bale05}. Here, the velocity of the slow solar wind is $V_{sw}\simeq350$~km/s, and $\beta_p\simeq5$. Figure~\ref{fig:spec}(b) gives the power spectral density $PSD(f)$ of $B_x(t)$, using (i) the Fourier transform,  and (ii) the Morlet wavelet transform \citep{farge92,torrence98,eriksson98}. The latter spectrum is defined as 
\begin{equation}\label{eq:PSD}
PSD_x[\textrm{nT}^2/\textrm{Hz}]=\frac{2\delta
t}{N}\sum_{j=0}^{N-1}|\mathcal{W}_x(\tau,t_j)|^2 \ ,
\end{equation}
where
\begin{equation}\label{eq:wt}
\mathcal{W}_x(\tau,t) = \sum_{j=0}^{N-1} B_x(t_j) \; \psi \left( \frac{t_j-
t}{\tau} \right)
\end{equation}
is the wavelet transform of $B_{x}$. We assume that the latter is regularly sampled at $t_j = t_0 + j \delta t$, with $j=0,...,N-1$. Here, the dilation parameter $\tau$ sets the time scale, while 
\begin{equation}
\psi (u)= \pi^{-\frac{1}{4}} e^{-j\omega_0 u} e^{-\frac{1}{2} u^2}
\end{equation}
is the unnormalised Morlet wavelet. Setting $\omega_0=2 \pi$ allows the characteristic scale to be inferred from  the angular frequency directly with $\tau = 2 \pi / \omega$.
Wavelet transforms allow us to resolve a non-stationary signal in time and scale (or frequency). Moreover, the analysing time window is scale-dependent and therefore better tailored to each scale. This is in contrast to the windowed Fourier transform, which keeps the same analysing window at all scales. Below (in Sect.~\ref{sec:struc}), we shall exploit the time-scale resolution of wavelets; here we just use them for power spectral estimation purposes.

In Figure~\ref{fig:spec}(b) the wavelet spectrum is multiplied by a factor 10 for ease of visualisation. Strong fluctuations in the Fourier spectrum hinder the visualisation of linear slopes. The wavelet spectrum, in contrast, is smooth and  clearly exhibits a power law with a ${-5/3}$ slope, within the $\Delta f=[10^{-3},0.2]$~Hz frequency range. 

To verify this spectral law and help better determine the frequency range, it is useful to compensate the observed spectrum with an inverse of the observed law: if the spectrum follows indeed a power law $\sim f^{-5/3}$, the compensation by $f^{5/3}$--function should give a constant.  Figure~\ref{fig:spec}(c) gives the compensated spectrum with $f^{5/3}$ function (solid line). The resulting function is flat over the $\Delta f=[10^{-3},0.2]$~Hz frequency range (bounded by the vertical dotted lines in   panels (b)-(d)). This compensated spectrum shows the ion break point at 0.3~Hz, which was not so immediate from the previous spectrum.

To check the comparison with a Kolmogorov spectrum, we compensate the observed wavelet spectrum with the inverse of the Iroshnikov-Kraichnan (IK) law $f^{3/2}$ (dashed-dotted line). The result is not constant, indicating that the IK--model is not correct. Note that $5/3-0.1(5/3)=3/2$ and so, if the precision on the spectral index is close or superior to  $10\%$, one would not be able to distinguish between these two scalings.  Figure~\ref{fig:spec}(d) shows the Fourier spectrum compensated by the inverse of the Kolmogorov law. Strong fluctuations, especially for $f>10^{-2}$~Hz, preclude the robust estimation of the spectral index. We conclude that the wavelet spectra help us in determining the spectral index and the frequency range of interest, where the power law is valid. 

Numerous examples of the use of the wavelet spectra can be found in geospace plasmas \citep{eriksson98,bale05,alexandrova08c,alexandrova09,kiyani13} and in laboratory plasmas \citep{stroth04,vandenberg04,zweben07}.

The practical problem of power law identification and spectral index estimation is of considerable importance and has also led to a large number of erroneous interpretations. Quite often, spectral indices are derived from ranges of frequencies (or wavenumbers) that span less than a decade, which is extremely risky. Another frequent mistake is the fitting of power laws to spectra that show no compelling evidence for a linear slope in logarithmic coordinates. Rigorous frameworks such as the maximal likelihood approach, as discussed by \citet{clauset09}, ought to be a standard for any power law study. Another issue is the unbiased estimation of spectral indices, for which discrete wavelet transforms are preferable to continuous wavelet (\textit{e.g.} Morlet) or Fourier transforms because they are unbiased \citep{abry95}. Undecimated discrete wavelet transforms (also known as the \textit{\`a trous} method) are of particular interest because they provide an orthonormal scale-dependent basis while keeping the timing information that is important for visualisation \citep{kiyani13}.

Power spectra and their spectral indices have received much attention so far because they can be conveniently estimated from time series or computed directly by means of spectral analysers. The spectrum, however, only describes a second order moment of the turbulent wavefield and as such gives a very incomplete picture of the system. Higher order moments and their properties will be addressed in the next Section.    

%%%%%%%%%%%%%%%%%%%%%%%%%%%

\subsection{How to measure departure from self-similarity: PDFs and structure functions}
\label{sec:strucf}

The most complete tool for describing the statistical properties of the field fluctuations at different scales, including departure from self-similarity, is the probability density function (PDF) $\mathcal{P}(y)$ of the random variable $y$, defined as $\mathcal{P}(y) \; du = P(u<y < u+du)$, where $P$ stands for the probability. 

\begin{figure}[!htb] 
\centering
\includegraphics[width=0.80\textwidth]{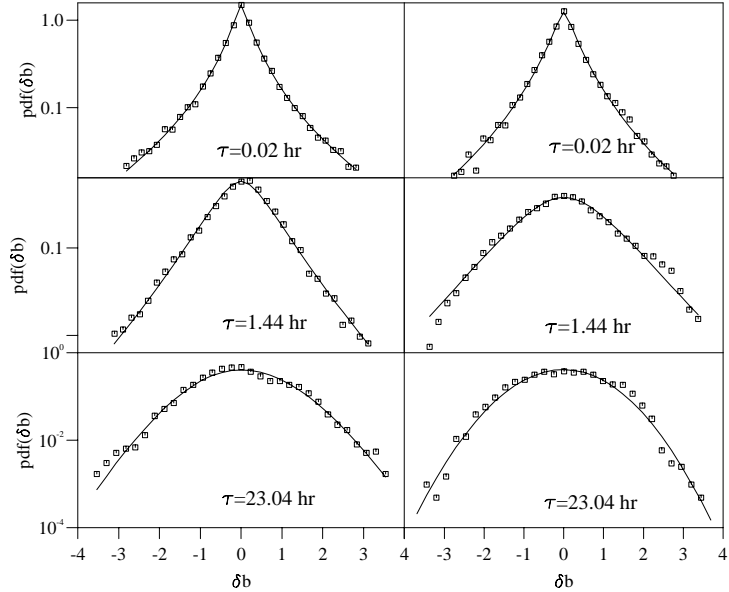}
\caption{Deviation of the PDFs from Gaussian statistics with scale: signature of intermittency in the inertial scale of the solar wind magnetic field \citep{sorriso-valvo99}. Left panels are for fast solar
wind, and right panels for slow solar wind. \label{fig:intermit}
 }
\end{figure}

For a given turbulent record,  fluctuations at different time scales $\tau$ can be approximated by increments, defined as 
\begin{equation}
\Delta y_{\tau}(t)=y(t+\tau)-y(t) \ .
\label{eq:increments}
\end{equation}
Statistical properties of the turbulent record at different scales $\tau$ can be investigated  by plotting the probability density $\mathcal{P}(\Delta y_{\tau})$. Figure~\ref{fig:intermit} shows a typical behaviour of the PDFs of magnetic fluctuations in the solar wind inertial range \citep{sorriso-valvo99,sorriso-valvo01,carbone04,hnat2003,leubner05,kiyani07}. The same properties are observed in laboratory plasmas \citep{carbone2000,sorriso-valvo01,antar01,carbone04,hnat08}. 
Statistical self-similarity implies that the PDF at scale $\tau$ can be collapsed onto a unique PDF $\mathcal{P}_0$ by following transformation
\begin{equation}
\mathcal{P}(\Delta y_{\tau}) = \tau^{-H} \mathcal{P}_0(\Delta y) \ .
\label{eq:selfsim}
\end{equation}
where $H$ is the Hurst exponent. Equation~(\ref{eq:selfsim}) implies that the increments are self-affine, namely $\Delta y_{a\tau} = a^H \Delta y_{\tau}$, where $a$ is a scaling parameter.

The departure of the PDFs from a Gaussian with decreasing time scale is usually a signature of intermittency, and indicates that turbulent fluctuations are not self-similar at different scales.

The tails of the PDF are of particular interest, because the distribution of rare events is indicative of the nature of underlying physical process. However, the practical assessment of such tails is a delicate task, and so moments of the PDF often receive more interest than the PDF itself. The moments of $\mathcal{P}(|\Delta y_{\tau}|)$ are called structure functions and can be estimated directly from the time series as 
   \begin{equation}
   S_p(\tau)= \int_{-\infty}^{+\infty} \mathcal{P}(|\Delta y_{\tau}|) \; |\Delta y_{\tau}(t)|^p \, dt = \langle |\Delta y_{\tau}(t)|^p \rangle \, ,
   \label{Eq-Self-Struc}
   \end{equation}
where $\langle \cdots \rangle$ denotes ensemble averaging.
Equation~\ref{eq:selfsim} implies that the structure functions should scale with $\tau$ as
\begin{equation}
  S_p(\tau)\propto \tau^{\zeta_p} .
\end{equation}
For statistically self-similar processes, the scaling exponents $\zeta_p$ are a linear function of the order $p$; deviations from this linear behaviour can thus be used as a quantitative measure of departure from self-similarity. There is considerable experimental evidence that turbulent flows deviate from this behaviour \citep{frisch95}.
   
Solar wind and laboratory data have been extensively studied by structure function analysis, showing the presence of intermittency \citep{carbone94,tu95,carbone2000,antar01,bruno05,matthaeus11}.
The evaluation of structure functions is straightforward, but there are pitfalls. The main danger is the increasing sensitivity of structure functions to rare and large events when the order $p$ increases, until finite sample effects completely dominate. This often goes unnoticed as the structure function increases smoothly with order. As a rule of thumb, it is considered safe to compute structure functions up to order 
 \begin{equation}
p_{max}=\log N -1 \ ,
   \end{equation}
where $N$ is the number of samples in the dataset. A more detailed check relies on the convergence of the integrand in Eq.~(\ref{Eq-Self-Struc}) \citep{ddw04}. Recursive convergence techniques have also been applied to solar wind data \citep{chapman09}. 

Another common measure of intermittency is the normalised fourth order moment of PDF, also known as  flatness \citep[\textit{e.g.},][]{bruno05} 
   \begin{equation}
      F(\tau)=S_4(\tau)/S_2^2(\tau) \ .
      \label{eq:flatness}
   \end{equation}
For Gaussian fluctuations 
%of a random variable $y$ at a fixed scale $\tau$, $P(y)=\exp{-y^2/2}/\sqrt{2\pi}$,
%$$S_4=\int y^4 P(y) =  3\int y^2 P(y) = 3 S_2 $$ therefore, 
$F(\tau)=3$, so that a scale-dependent departure from this value can be used as an indication for intermittency.  

Note that the increment defined in Eq.~(\ref{eq:increments}) is merely the convolution of the turbulent wavefield $y(t)$    
\begin{equation}
      \Delta y_{\tau}(t) = \int y(t') W_{\tau}(t-t') \; dt' \ ,
      \label{eq:increments2}
   \end{equation}
with the coarsest wavelet one could imagine, namely $W_{\tau}(t) = \delta(t+\tau)-\delta(t)$. The statistical properties of turbulent eddies can also be captured with increments that involve more elaborate wavelets (in terms of continuity, number of vanishing moments, \textit{etc.}). This generalisation establishes an connection with a much broader framework wherein wavelet coefficients and their higher order moments are the prime quantities of interest \citep{muzy93,farge04}. Interestingly, this framework has intimate connections to the thermodynamic formalism.

Finally, in those cases where the estimation of the structure functions is handicapped by a short inertial range, the Extended Self-Similarity (ESS) technique \citep{benzi93} can help extract relative scaling exponents \citep{carbone96}. Let us stress again, however, the importance of having long records for properly accessing high order moments, in particular with heavy-tailed processes \citep{kiyani09b}.

%%%%%%%%%%%%%%%%%%%%%

\subsection{Coherent structure identification}
\label{sec:struc}

The origin of the intermittency in plasmas is still an open question and is usually ascribed to the presence of coherent structures whose typical lifetime exceeds that of incoherent fluctuations in the background. Typical examples are current sheets, shocks and vortices in space plasmas \citep{veltri99a,veltri99b,sorriso-valvo99,mangeney01,alexandrova06,alexandrova08a,alexandrova08d,carbone2000,greco09b,greco12}, and blobs, vortices, clumps, avaloids, and similar structures in laboratory plasmas \citep{zweben85,huld91,antar03,krasheninnikov08,fujisawa10}. 

A simple definition of a coherent structure is a structure whose phases are coupled for a finite range of scales, or at all measured scales. For example, coherent structures in fluid turbulence often are localized vortex filaments whose length is of the order of the energy injection scale $L$, and whose cross-section is of the order of the dissipation scale $\ell_d$ \citep{frisch95}; in Fourier space, these structures occupy all scales from $L$ to $\ell_d$. Although there is no universal definition of what a coherent structure is, phase coherence is an essential ingredient, which can be tested, if not observed in Fourier space. One can indeed build a test statistic for the existence of coherent structures by randomising the phases of the Fourier transform, while keeping the amplitudes unchanged. This operation preserves the power spectral density and second order quantities such as the autocorrelation function, but destroys phase synchronisations.  This surrogate and its statistics can be compared to the observed signal and its statistics.   
Surrogates are widely used in dynamical systems \citep{schreiber00b} to test the null hypothesis against a linear, Gaussian, and stationary stochastic process; some applications to plasmas have been reported as well \citep{hada03,sahraoui08,chian09}. 

While Fourier transforms are routinely used for making surrogates and testing phase coherence, it often makes more sense to detect phase coherence locally in time. Multiresolution (\textit{i.e.} wavelet) techniques again stand out as the most powerful tool for dealing with such problems. Here, the quantity of interest is the energy distribution of the time series in scale $\tau$ \textit{and} in time $t$; notice that the latter had been ignored in the determination of frequency spectra in Sect.~(\ref{sec:spectra}).  

We already defined the wavelet transform in Eq.~(\ref{eq:wt}). Ideally, the mother wavelet function should be  tuned to the kind of coherent structure one wishes to analyse. For example, Haar (or 0th order Daubechies) wavelets
\begin{equation}
 \psi(t) =\left\{ 
 \begin{array}{rl} 
 		0, & \textrm{ for $|t| \ge 1/2$ } \\
 		1, &\textrm{ for $0\le t<1/2$} \\ 
 		-1, & \textrm{ for $-1/2<t<0$} 
 \end{array}\right. \ .
\end{equation}
are appropriate for analysing  sharp discontinuities (non-differentiable functions). First order Daubechies  are better suited for handling smoother structures, whose first derivative exists but not the second one, \textit{etc.} Haar wavelets have been used in various contexts to identify shocks and current sheets in the solar wind inertial range of scales \citep{veltri99a}, to identify events in fusion plasmas \citep{vega08}, and to de-noise plasma simulation data \citep{nguyen10}.

Note that the Haar wavelet transform is equivalent to a set of increments at scales defined as powers of $2$ of the smallest scale: $\tau_m=2^m\tau_0$. Using these increments, a different method based on variance threshold, referred to as Partial Variance of Increments (PVI), was recently used by Greco and al. to identify magnetic discontinuities in numerical simulations and in solar wind time series \citep{greco08,greco09,greco09b,greco12}.
 
\begin{figure}[!htb] 
\centering
\includegraphics[scale=1.00]{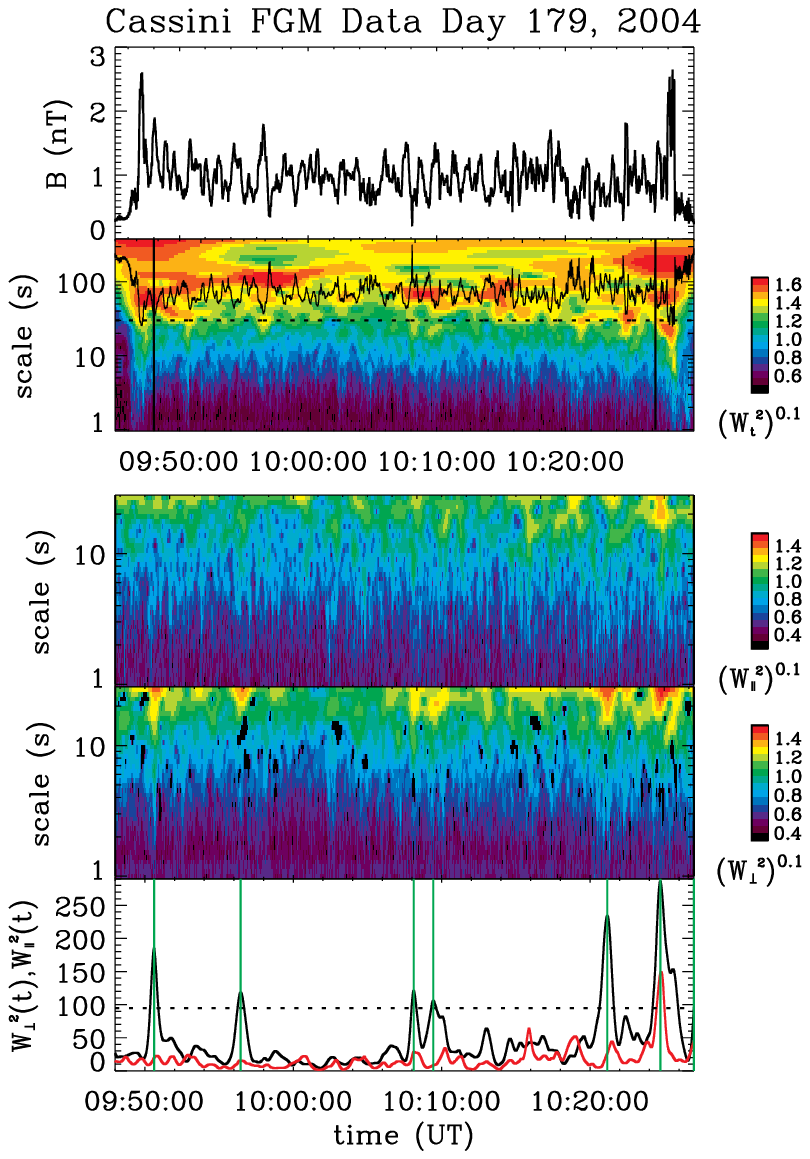}
\hspace*{8mm}
\includegraphics[scale=0.46]{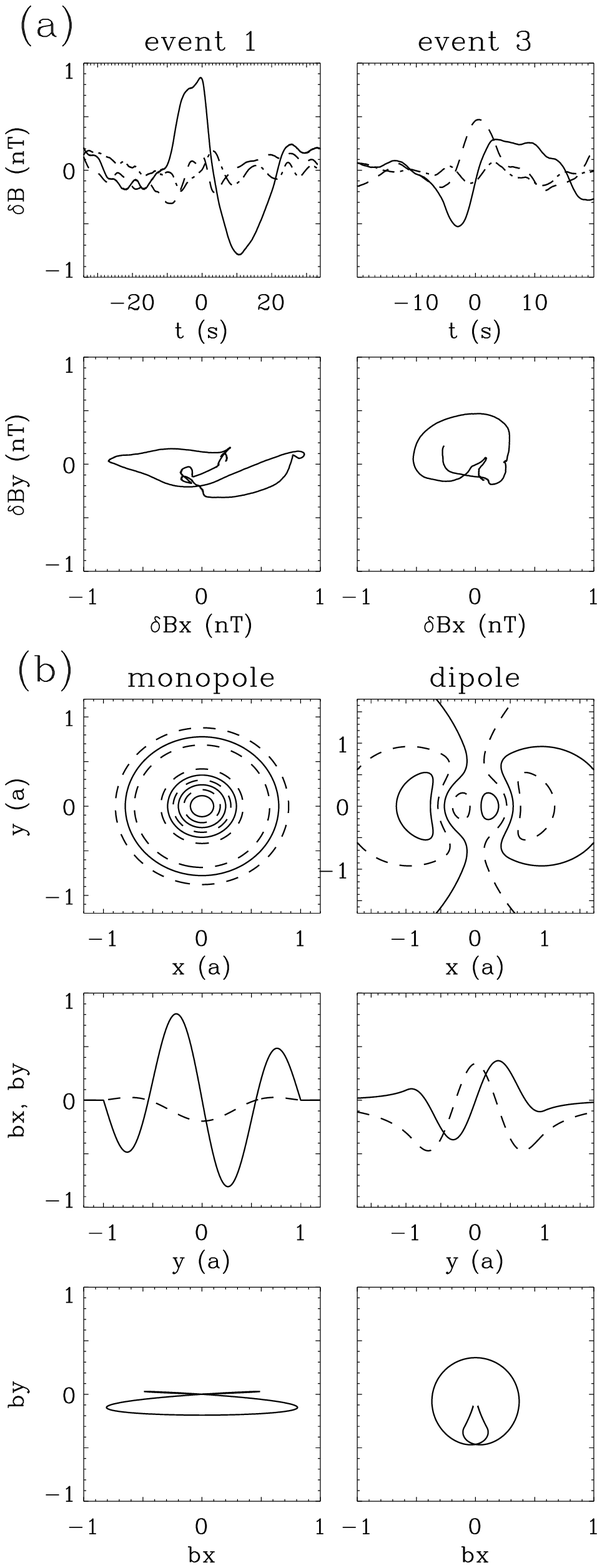}
\caption{Left, from the top to the bottom: Magnetic field intensity; Morlet wavelet scalogram, which shows how the total energy of magnetic fluctuations (in colour) is distributed in time-scale plane; below we separately the energy in compressible fluctuations $W_{\|}^2(t,\tau)$ and the distribution of the energy of Alfv\'enic  fluctuations $W_{\perp}^2(t,\tau)$; the bottom panel gives the integration of $W_{\|}^2(t,\tau)$ (red line) and $W_{\perp}^2(t,\tau)$ (black line) over scales. The peaks of the energy of Alfv\'enic fluctuations (see green vertical lines) correspond to coherent structures, two of them are shown on the right: (a) 3 components of magnetic fluctuations in the local magnetic field frame (top), polarisation of the fluctuations in the plane perpendicular to the magnetic field (2nd line); (b) comparison with models of monopole and dipole Alfv\'en vortices. Here, $a$ is the vortex radius. For more details, see \citep{alexandrova08d}. \label{fig:saturn}
}
\end{figure}

Other wavelets have been used as well, such as Morlet wavelets for detecting magnetic Alfv\'en vortices in the Earth's and in Saturn's magnetosheath.  Figure~\ref{fig:saturn} illustrates how the structures were identified at Saturn during the Cassini orbit insertion in 2004 \citep{alexandrova08d}.

A popular measure of coherent structures is the Local Intermittency Measure (LIM). The LIM does not truly measure intermittency or coherent structures, but rather local concentrations of energy in the spectral domain, \citep[\textit{e.g.},][]{bruno05}. If $W(t,\tau)$ is the wavelet transform of a process at time $t$ and scale $\tau$, then 
\begin{equation}
\textrm{LIM}(t,\tau) = \frac{|W(t,\tau)^2|}{\langle |W(t,\tau)|^2 \rangle_t}
\end{equation}
Any value significantly exceeding 1 is an indication that the wave field has local concentration of energy. This typically occurs with bursts of wave activity, such as whistler waves at shock fronts, or with discontinuities. Smooth coherent structures such as blobs may easily be missed by the LIM, which, by definition, performs best with structures whose energy is concentrated in time and scale. The LIM is easier to interpret when using complex wavelets, since it then relies on the envelope of the wavelet transform and is not affected by the phase of the latter. One can similarly define a LIM that is based on fourth order moments, such as $\textrm{LIM}^2(t,\tau) = |W(t,\tau)^4| / \langle |W(t,\tau)|^4 \rangle_t$.

Let us stress that Haar wavelets, and low order wavelets in general, formally are relevant only  for  sharp discontinuities. These wavelets cannot properly characterise more continuous structures and fail to provide the spectral index of time series whose power law is steeper than a given threshold value \citep{farge06}. For that reason, higher order wavelets such as 4th or 6th order Daubechies should systematically be preferred when it comes to analysing coherent structures. Numerous other examples of coherent structure identification and extraction can be found in neutral fluid simulations \citep{farge92,farge01}. 

So far we only discussed methods that are applicable to single time series. In Sect.~\ref{sec:spatiotemporal} we shall elaborate on spatio-temporal observations, which are more relevant for analysing coherent structures, but also more challenging for the observer.

%%%%%%%%%%%%%%%%%%%%%%%%%%%%%%%%%%%%%%%%%%%%%%%%%%%%%%%%%%%%%%%%%%%%%%%%%%%%%%%%%%%%%%%%%%%%%%%%%%%%%%%%%%%%%%%%%%

\section{Spatio-temporal observations of turbulent wavefields}
\label{sec:spatiotemporal}

In many observations, the temporal and spatial dimensions are intimately mixed because of plasma motion, and also because the probes often are not at rest in the plasma frame. As a consequence, the distinction between spatial structure and temporal dynamics is often elusive, except in the case where at least two simultaneous observations of the same variable can be made. The different methodologies that have been developed for that purpose depend very much on the number of simultaneous observations, which will be here our guiding thread. Here we address mostly wave-like structures but also structures of arbitrary shape.

%%%%%%%%%%%%%%%%%%%%%%%%%%%%%%%%

\subsection{Two-point measurements}

Single-point measurements are formally inappropriate for disentangling space and time in plasmas. However, in the particular case where the plasma flow is steady, dispersionless,  and fast with respect to its fluctuations, Taylor's hypothesis allows to infer spatial structures from temporal variations. This hypothesis is frequently used in the solar wind to infer wavenumber spectra and spatial structure functions from single time series, see for example \citet{horbury11} and \citet{marsch97}. 

Two closely spaced measurements (as compared to the characteristic spatial scales of the medium) give access to a wealth of new information. The prime quantity of interest is the joint frequency-wavenumber spectrum $S(k,f)$, which is the key to the identification of plasma waves. The first step towards the extraction of that spectrum is the estimation of the cross spectrum
\begin{equation}
S_{XY}(f) = \langle Y(f) \, X^*(f) \rangle = | S_{XY}(f) | \; e^{j \phi_{XY}(f)} \ ,
\end{equation}
where $X(f)$ and $Y(f)$ are respectively the Fourier transforms of the two simultaneously measured scalar observables $x(t)$ and $y(t)$. The phase $\phi_{XY}(f)$ of this complex quantity is related to the wavenumber $\mathbf{k}$ projected along the separation vector $\mathbf{d}$ of the two measurements by 
\begin{equation}
\mathbf{k} \cdot \mathbf{d} = \phi_{XY}(f) \ .
\label{eq:wavenumber}
\end{equation}
This property has been widely used to infer wavenumbers in simple configurations wherein the direction of propagation is known; the latter is usually determined by minimum variance analysis \citep{sonnerup98}. An early example is the identification of whistler waves in the solar wind using two the nearby ISEE spacecraft \citep{hoppe80}.

Classical cross-spectral analysis, however, has several limitations. First, it cannot properly distinguish multiple waves that have different dispersion relations. To overcome that  limitation, \citet{beall82} proposed to use instead the local joint frequency-wavenumber spectrum, defined as
\begin{equation}
S_{L}(k,f) =  \sum_i   \frac{1}{2} \left[ X_i(f) X_i^*(f) + Y_i(f)Y_i^*(f) \right] \delta\big( k_i(f) - k \big)  \ ,
\label{eq:beall}
\end{equation}
where the local wavenumber $k_i(f) = | \mathbf{k}_i(f)|$ is estimated from the cross-spectrum, for different ensembles that are indexed by $i$.  According to this equation, the frequency-wavenumber spectrum is obtained by unfolding the frequency spectrum, conditioned by the wavenumber defined in Eq.~(\ref{eq:wavenumber}). This approach has been widely used in laboratory plasmas, and in particular for drift turbulence studies, using Langmuir probe pairs \citep{poli06,tynan09}.

There is a second limitation, however, which has often been overlooked: all these Fourier-based techniques are formally applicable only to plane waves with an infinite spatial extent. Such conditions are rarely met in turbulent plasmas, which are often non-stationary and rather consist of a superposition of wave-packets. This is even more so in strong turbulence, in which the size of the wave packets becomes comparable to their characteristic period or wavelength. The generalisation of Eq.~(\ref{eq:beall}) to multiscale methods allows to overcome this limitation \citep{ddw95}. The idea simply consists in replacing the Fourier transform in Eq.~(\ref{eq:beall}) by a continuous wavelet transform.

\begin{figure}[!htb]     \begin{center}
    \includegraphics[width=0.32\textwidth]{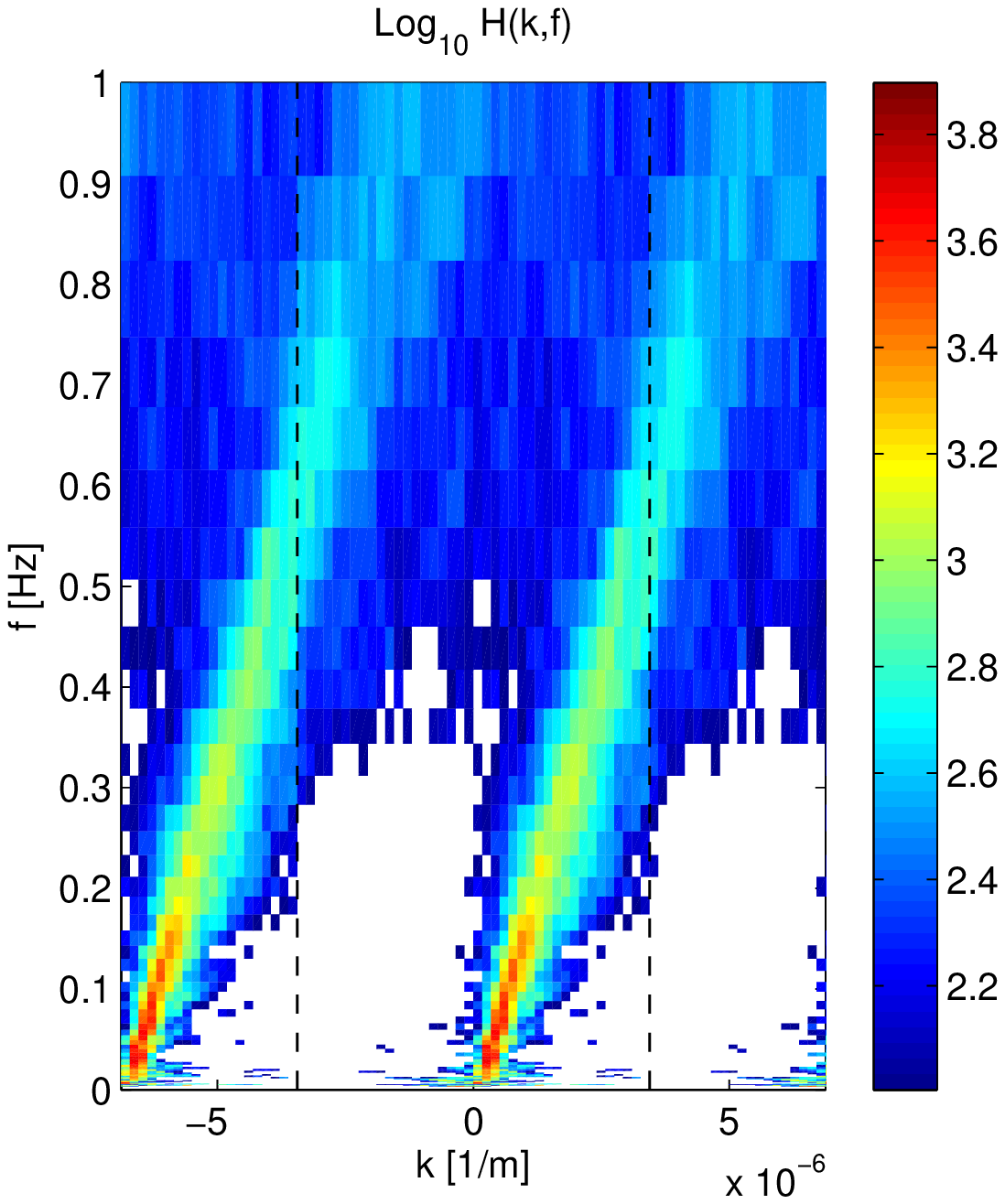}
    \includegraphics[width=0.33\textwidth]{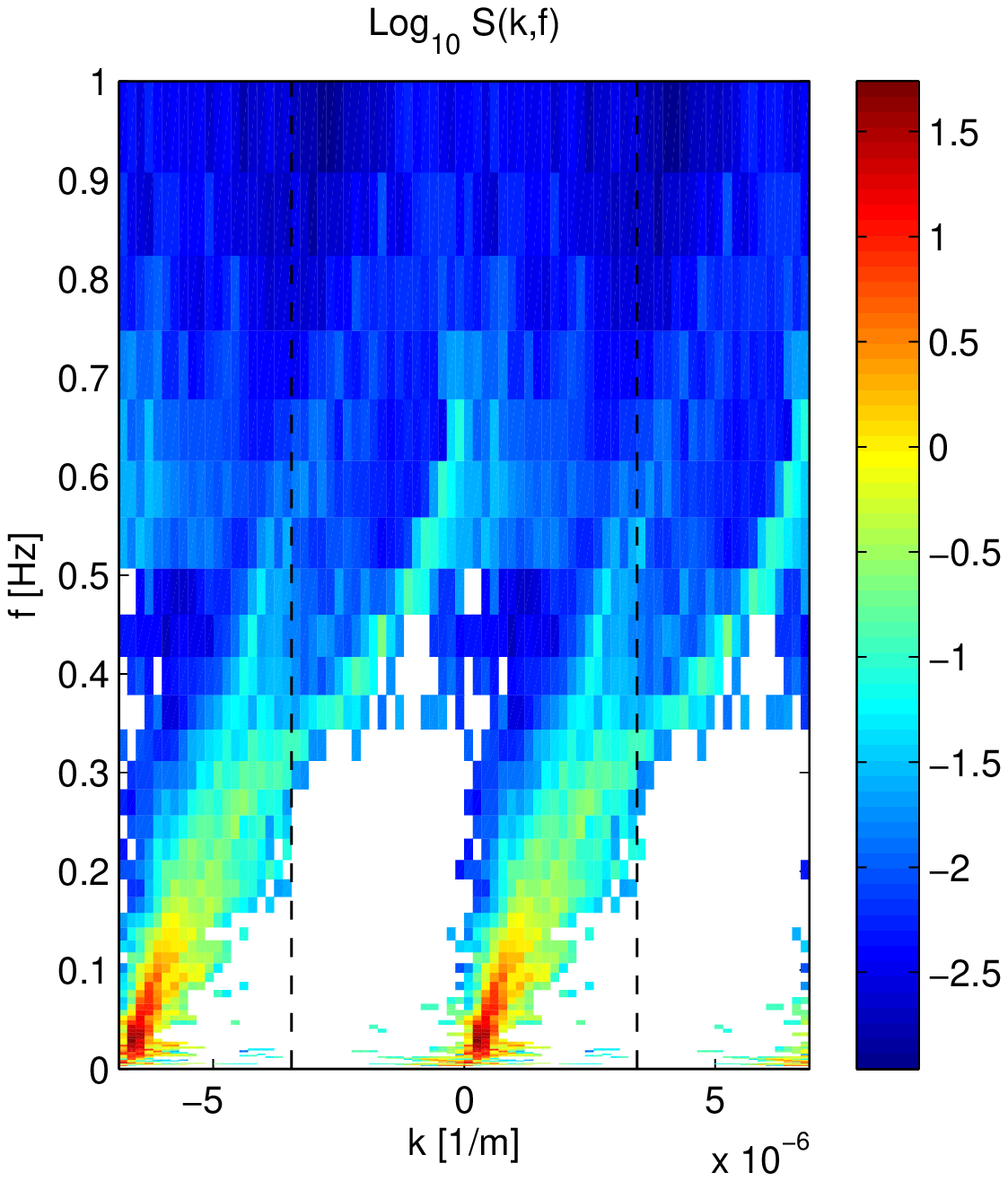}
    \includegraphics[width=0.32\textwidth]{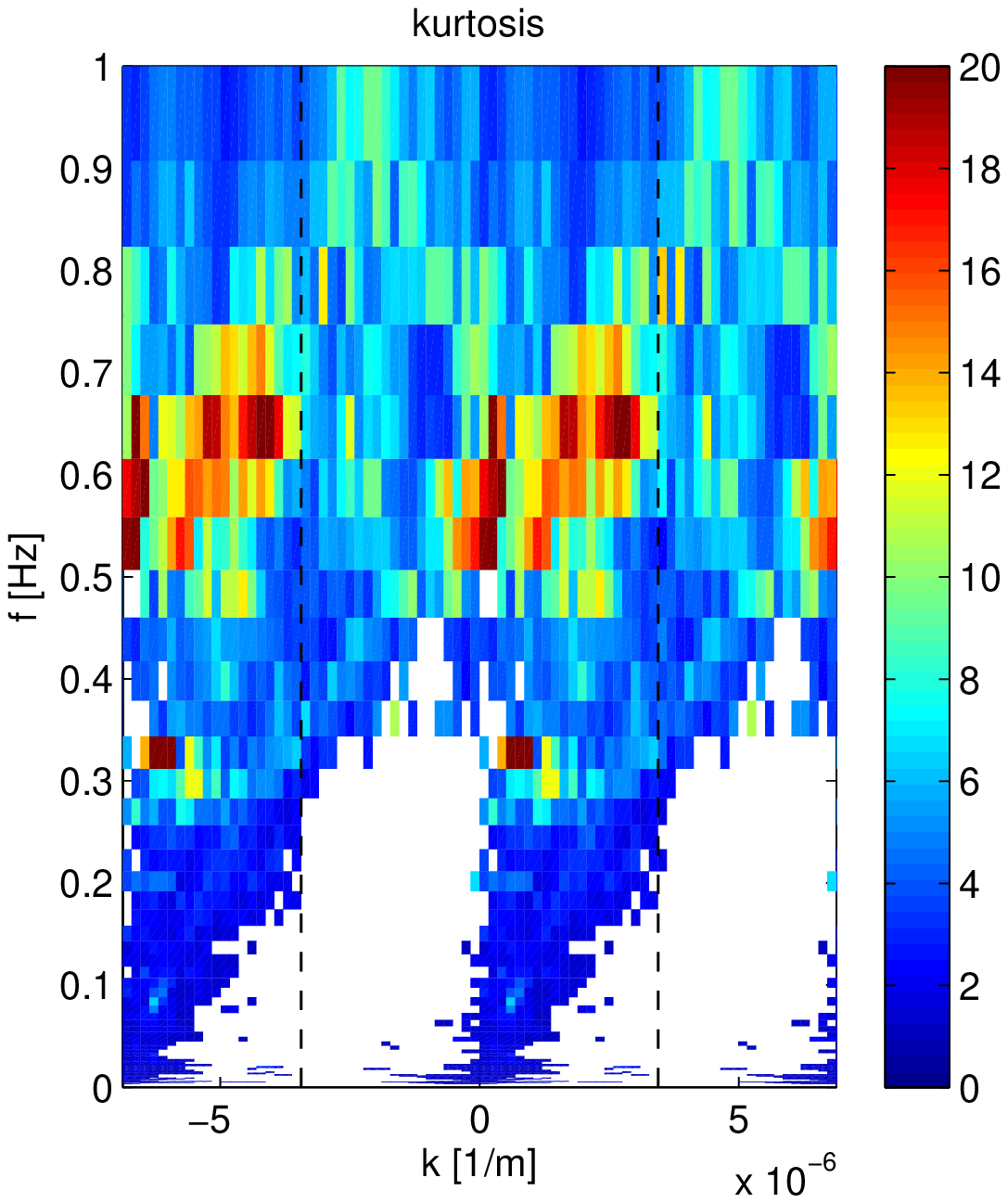}
    \end{center}
    \caption{Wavefield properties as a function of frequency and wavenumber, inferred from two-point magnetic field measurements in the foreshock region, from the AMPTE-UKS and AMPTE-IRM spacecraft. Details are given in \citep{ddw95}. The left plot displays the number of wave packets, regardless of their amplitude; the middle plot shows the local frequency-wavenumber spectrum $S_{L}(k,f)$ estimated with wavelets; the right plot shows the flatness, as a signature of temporal burstiness. The dashed rectangle represents the principal domain, which is bounded by the spatial Nyquist number, and has been artificially unfolded. The only physically meaningful branch starts at $(f=0, k=0)$. \label{fig_kfspect}
}
\end{figure}

Figure~\ref{fig_kfspect} shows a typical example taken from the solar wind magnetically connected to the Earth's bow shock (i.e. foreshock region). 
% Thierry, Fig. 4 represents the same data as in your 1995 GRL? In the foreshock and not in the solar wind, right?
One single linear branch is observed in the dispersion relation when all wave packets are considered, regardless of their amplitude. According to this picture, the apparent motion amounts to simple dispersionless advection. However, the local frequency-wavenumber spectrum, in which each wave packet is now weighted by its amplitude, clearly reveals the existence of two branches, with different group velocities. The slowest branch (in the satellite rest frame) is associated with dispersive whistler wave packets whose bursty nature is attested by their high flatness, see Eq.~(\ref{eq:flatness}). Surprisingly, the use of wavelets for estimating dispersion relations is well accepted in geospace plasmas \citep[\textit{e.g.}][]{eriksson98,lundberg12}, whereas they are still exceptional in laboratory plasmas \citep{lazurenko08}. This example shows how long it may sometimes take for a ideas to spread.

The concept of cross-spectrum can be generalised to higher order spectra in order to describe nonlinear interactions, see for example \citep{kim79,kravtchenko95,ddw03}. Higher order spectra are relevant  for weak turbulence, because nonlinearities can be adequately described by three-wave and four-wave interactions. Interestingly, in the Hamiltonian framework for weak turbulence \citep{musher95}, there is a direct connection between the physical model and higher order spectra. Langmuir turbulence is one of the few examples in which empirical statistical quantities have such an immediate physical meaning. 

The seminal work of \citet{ritz86}, who showed how to derive  spectral energy transfers from higher order spectra using double probe measurements, opened the way to several applications. Indeed, energy transfers give deep insight into the nonlinear interactions. They allow, for example, to determine whether coherent structures in a turbulent wavefield are dynamically evolving or are static remnants of a nonlinear process that took place earlier. There have been surprisingly few applications so far; some  can be found in density fluctuations in tokamak edge turbulence  \citep{kim97,zweben07}, in magnetic field fluctuations in the solar wind  upstream the Earth's quasi-paralle bow-shock %, in which Short Large Amplitude Magnetic Structures (SLAMS) play a leading role. 
 \citep{ddw99}, and in Langmuir turbulence simulations \citep{soucek03}. Let us stress that they require long records, stationarity, and need careful validation to avoid misinterpretation. This may explain why they have been mostly applied to laboratory plasmas.

For fully developed turbulence, or when the spectral description is less appropriate,  other  techniques may be more relevant. The most elementary one is the cross-correlation function, which informs about temporal lags and about the decorrelation time. More interesting is the conditional averaging technique, also know as superposed epoch analysis in space science. The technique consists in taking a snapshot of the measurements whenever a reference signal exceeds a threshold. Averaging these events allows the coherent patterns, if there are any, to emerge from the sea random fluctuations. Conditional averaging has been widely used to identify and characterise spatially coherent structures (often referred to as blobs or clumps) in electrostatic turbulence \citep{johnsen87,windisch06,diallo08}. In geospace plasmas, conditional sampling has been more frequently used for identifying the shape of transients \citep{baker86,borovsky10}. Laboratory applications often offer the additional advantage of allowing the probes to be moved, so that a full range of spatial scales can be investigated with two probes only, just by repeating the experiment under the same conditions \citep[\textit{e.g.},][]{furno08,fattorini12}. These techniques, combined with smart detection criteria, are gaining again a lot of interest.

%%%%%%%%%%%%%%%%%%%%%%%%%%%%%%%%

\subsection{Three-point measurements and more}

The transition from two-point measurements to three-point (and more) is relatively minor in terms of gain of information. Many multi-probe studies actually are made by processing the probe signals pairwise. A beautiful exception in space research is given by the 4 satellites from the CLUSTER mission, which form a tetrahedron, thereby providing access to full 3D resolution in space.  With such a configuration, the determination of gradients, of the three components of wave-vectors, and normals to natural boundaries becomes possible. This has led to important improvements in the study of microphysical processes in space plasmas, both qualitatively and quantitatively. Most of these techniques and the first results obtained with them are detailed in the two books by \citet{paschmann00,paschmann08}, which we refer the reader to. 

Some of the physical properties that have been inferred from CLUSTER observations, are: the estimation of the current density from the curl of the magnetic field \citep{haaland04b}, the analysis of the geometrical structure of discontinuities such as shocks \citep{shen03}, the identification of a co-moving deHoffmann-Teller frame using wavefield measurements \citep{khotyaintsev04}, and Alfv\'en vortices in the Earth's environment \citep{sundkvist05,alexandrova06}. 
Several successful attempts have been made for reconstructing the three-dimensional spatial spectrum of turbulence over at least one decade of scales  \citep{tjulin05,sahraoui06,sahraoui11}. These reconstructions are based on the $k$-filtering technique \citep{pincon91}, which allows to overcome the limitations of the Taylor hypothesis.  Many of these studies have been instrumental in unveiling key physical processes such as reconnection events at the Earth's bow shock \citep{retino07} and at providing a microscopic look at the processes allowing particles from the Sun direct entry into the magnetosphere \citep{hwang12}. Cross-Scale, the follow-on of CLUSTER, was designed to probe both ion and electron scales, and would thereby have opened new perspectives in multipoint data analysis \citep{dunlop11}. To finish, let us mention a particular case wherein these techniques have been adapted to three-point measurements \citep{vogt09}.

Here we would like to stress again the importance of understanding the limitations behind spectral techniques such $k$-filtering, which assume a superposition of uncorrelated plane waves with random phases. These assumptions are rarely met in practice and great care must be taken in analysing cases such as spatially coherent structures that are convected with the flow.

Another conceptually interesting  approach involves multipoint observations of systems that exhibit ``global'' modes. The latter encompass here anything going from traveling waves to perturbations affecting the system globally or partly. In such spatio-temporal systems, it often makes sense to look for separable solutions of the observable $f(\mathbf{x},t)$:
\begin{equation}
f(\mathbf{x},t) = \sum_k \phi_k(t) \; \psi_k(\mathbf{x}) \ .
\end{equation}
This decomposition can be unique if constraints are imposed on the modes $\phi_k(t)$ and  $\psi_k(\mathbf{x})$. Various techniques have been developed in what has become a highly multidisciplinary field of research that is also known as \textit{blind source separation}: the Singular Value Decomposition or Biorthogonal Decomposition are used when the modes are orthonormal, Independent Component Analysis, when they are independent, \textit{etc.} \citep{ddw11}. In all these techniques, the heuristic idea is to concentrate the salient features of the spatio-temporal wavefield in the smallest possible number of modes. These techniques excel in separating oscillations with different poloidal mode numbers in laboratory devices \citep{nardone92,ddw94,madon96,kim99,niedner99,dinklage00} and, more generally, in extracting variations that are spatially coherent \citep{manini03}. Such variations may be travelling waves but also non wave-like structures. Mode decomposition techniques are also useful for handling inverse problems such as those occurring in plasma tomography \citep{anton96}. Applications to geospace plasmas have been slow to come because multichannel observations are so few. A notable exception is the analysis of solar  and stellar spectra, which are just another kind of bivariate data \citep{amblard08,christlieb02}.

There was initially some hope that such techniques might help identifying and extracting coherent structures from turbulent wavefields, as suggested by \citet{benkadda94}. Few attempts turned out to successful. Indeed, techniques such as the Singular Value Decomposition rely on second order moments only and so formally cannot properly extract structures whose main hallmark is their departure from a Gaussian PDF. Multiscale techniques are better suited for that, as discussed, for example, by \citet{farge06} and by \citet{dippolito11}.

%%%%%%%%%%%%%%%%%%%%%%%%%%%%%%%%

\subsection{Imaging}
\label{sec:imaging}

Ideally, a complete characterisation of turbulence and meso-scale structures in plasmas, which we generally refer to as "imaging" in this Section, would require full spatio-temporal measurements of the relevant fields (i.e electron and ion density, temperature, plasma potential, \textit{etc.}) with adequate spatial and temporal resolution, and without perturbing the plasma. From the experimental point of view, this task is hampered by the intrinsic difficulty in diagnosing the region of interest with adequate temporal and spatial resolution. The seminal study by \citet{zweben85}, with a grid of $8 \times 8$ Langmuir probes, has remained unmatched for many years until new  techniques have enabled more quantitative information to be extracted from such data sets.

We will not address here the remote sensing of solar and heliospheric plasmas, whose breathtaking images have stimulated the development of a wide variety of techniques, some of which are very specific to the structures of interest, such as coronal loops, or the motion of granules in the photosphere. For some references, see  \cite{georgoulis05,meunier09,aschwanden10,aschwanden11c}. The next conceptual step will involve data cubes such as two dimensional (2D) images taken in different wavelengths by spectro-imagers or, more generally, hyperspectral images.

In recent years, progress has been made in basic plasma physics devices, where arrays comprising a large number (of the order of $10^2$) of electrostatic probes have been commonly introduced to typically measure ion saturation and floating potential signals \citep{katz2008,mueller06}. Two diagnostic issues are the quantitative interpretation of probe results, especially in magnetised plasmas where the theory of electrostatic probes is still incomplete, and the possible perturbing effects of the probes on the physics under study. Relatively few papers discuss these issues in detail, and normally some checks are made to verify that the probe does not affect the results from existing probes, but it is often not clear whether the structure and motion of blobs is unaffected by these probes.

\begin{figure}[!htb]
	\begin{center}
	\includegraphics[width=1.0\textwidth]{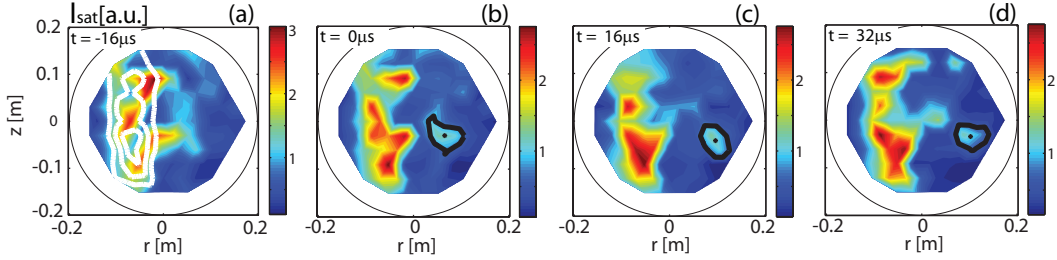}
	\caption{Example of fast imaging using 2D profiles of ion saturation current from a 2D Langmuir probe array on the TORPEX device. An intermittent turbulence-generated structure, \textit{i.e.} a blob (highlighted by the thick black line) is radially propagating in the poloidal plane of the device. \label{blob.fig}}
    \end{center}
\end{figure}

As an example, Figure~\ref{blob.fig} presents the application of 2D imaging to visualise intermittent meso-scale structures, or blobs, in the TORPEX device \citep{Fasoli2010}. Blob propagation is investigated using a 2D array of 86 Langmuir probes on a hexagonal pattern, the HEXTIP diagnostic described in detail, see \citep{mueller06}. HEXTIP measures ion saturation current signals with a temporal resolution of 4 $\mu$s, which is smaller than the typical auto-correlation time of the structures under study ($\approx 100$ $\mu$s), and 1.75 cm spatial resolution. The first step, common to all diagnostics and techniques providing real space temporal imaging of plasmas, is the definition of the meaning of structures. To this aim, pattern recognition techniques assuming more or less complicated definitions can be used. In this example, blobs are identified by adopting a threshold segmentation approach, which selects positive structures as regions where the ion saturation signal exceeds an appropriate value, usually expressed in terms of the standard deviation of the signals. The identified structures are bound by polygonal structures, shown in black contour in Figure~\ref{blob.fig}, which allow computing different observables linked to the structures. Examples of observables are the occupied area, the position of the center of mass of a structure, \textit{etc.} By considering subsequent time frames, structure trajectories can be identified together with splitting events, when a single structure breaks apart, and merging events, when multiple structures merge together. The observables defined for static structures can now be computed along the trajectory of the structure itself, and new observables can be defined, such as for example the structure speed. This process allows to visualise, \textit{i.e.} image, the structure dynamics in turbulent plasmas. Furthermore, the strength of the real space imaging is that it not only allows to identify individual structures, but also to build a statistical ensemble over which to compute the statistical/probabilistic properties of the turbulent field under investigation \citep{mueller06}. This can be used to verify in a statistical sense predictions from theory, such as for examples scaling laws \citep{Theiler2009}.

The use of insertable probes for a full imaging of the quantities of interest for plasma turbulence studies suffers from two interlinked limitations, even in low temperature plasma experiments. A small number of probes only marginally perturbs the plasma dynamical behaviour, but would lead to either only local measurements or to insufficient spatial resolution to investigate the turbulence multiple scales. To obtain adequate spatial resolution, one would need to insert a large number of measuring tips, which would significantly perturb the phenomena under investigation. In recent years, to overcome these limitations, fast framing cameras have been adopted to monitor light emission from magnetically confined plasmas and also in basic plasma physics devices. In particular the field of fast imaging has been pushed forward by the need to investigate intermittent transport events associated with the filaments in fusion and basic plasma physics devices \citep{kirk2006,benayed2009,grulke2006,Iraji10}. 

\begin{figure}[!htb]
	\begin{center}
	\includegraphics[width=1.0\textwidth]{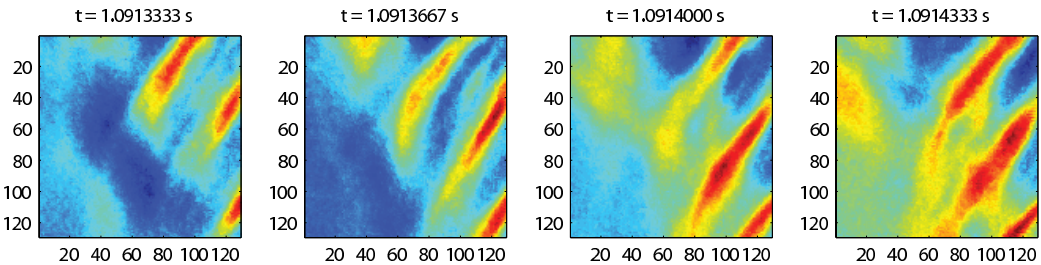}
	\caption{Example of filament evolution obtained using a fast framing camera at the edge of TCV tokamak. The axes are in units of pixel number. \label{filament_TCV.fig}
}
    \end{center}
\end{figure}

Fast cameras are today used to track filamentary structures with typical spatial scales $> 1$ mm and lifetime $>1$ $\mu$s in the scrape-off layer (SOL), and to compute their relevant quantities, such as average velocities and sizes. An example from the TCV tokamak \citep{Goodman03} is shown in Figure~\ref{filament_TCV.fig}. Fast framing cameras now offer $\approx 10^6$ pixels and acquisition frequencies up to $10^6$ images per second ($1$ $\mu$s time resolution), which are adequate to resolve most turbulent phenomena in magnetised plasmas. These large arrays of spatially distributed measurements open new possibilities of identifying turbulent structures in two-dimensional or even three-dimensional space and of following their evolution in time.

Two main difficulties are encountered when using fast cameras to image plasma turbulence. First, the signal is usually integrated along a line-of-sight resulting in multi-chord images, which need to be inverted by tomography, using more or less complicated techniques, to reconstruct the local plasma emissivity, see for example \citep{svensson08}. The line integration problem is common to other diagnostic techniques which can be generally classified as "imaging" techniques.
For example, measurements of line integrated electron density along satellite-to-ground path are obtained by using radio transmission from polar orbiting satellites \citep{Kersley97}. Tomographic inversion of these data allows mapping large portion of the ionosphere in a height versus latitude plane \citep{bust08}. The same approach can be applied to the solar corona \citep{butala10}. Tomographical reconstructions of heliospheric perturbations from observations of interplanetary scintillations (IPS) offer the ultimate example of how 3D structures can be reconstructed from line-of-sight observations \citep{jackson11}.

In laboratory devices, the line integration problem can be solved by using fast framing cameras together with gas puffing \citep{Agostini2009,Terry2003,Maqueda2010,zweben2010}, which result in atomic line emission from the locally injected neutral atoms. In these cases, precautions must be taken to avoid that the addition of a neutral source affects the local plasma parameters and turbulence features. The second problem is related to the intrinsic origin of the quantity measured by imaging, \textit{i.e.} optical plasma emission, which, in most cases, is a complex convolution of local density, temperature, and impurities concentration. This hampers a direct comparison of results from visible imaging with, for instance, numerical simulations of turbulent transport, unless these are coupled with radiation emission codes. The combination of forward modelling with MHD codes and direct IPS observations has been shown to enhance the reconstruction of solar wind structures \citep{hayashi03}.

Tomography inevitably leads to ill-posed problems, for which prior assumptions are important, and solutions are sensitive to the underlying assumptions. For such problems, the Bayesian framework \citep{vontoussaint11} has proven to be remarkably powerful and will surely inspire in the decade to come new techniques that can handle the imperfections of the data more properly.

%%%%%%%%%%%%%%%%%%

\section{Anomalous transport in plasmas}
\label{sec:anomalous}

The transport properties of suprathermal particles are important both for laboratory plasmas, where good confinement is essential in order to reach the goal of fusion, and in geospace plasmas, where understanding transport is important for predicting the arrival of solar energetic particles (SEPs) and for the acceleration processes like first order acceleration at shocks \citep{reames99,giacalone11,perri12a}. Here, an important issue that has challenged the analysis of microphysical processes is the characterisation of the transport processes.

In the classical regime of normal diffusion, the mean square displacement of particles along one dimension can be written as
\begin{equation}
\langle \Delta x^2 \rangle = 2D \, t \ ,
\label{eqzim1}
\end{equation}
where $D$ is the diffusion coefficient, and where $\langle \Delta x^2 \rangle$ grows linearly in time. However, in the last years it has become clear that anomalous transport regimes can be found in a large variety of physical systems, such that $\langle \Delta x^2 \rangle$ grows nonlinearly in time,
\begin{equation}
\langle \Delta x^2 \rangle = 2D_{\gamma} \, t^{\gamma} \ ,
\label{eqzim2}
\end{equation}
with $\gamma < 1$ in the case of subdiffusion, $1<\gamma<2$ in the case of superdiffusion, and with $\gamma = 2$ representing the ballistic regime \citep[\textit{e.g.},][]{zimbardo05,zimbardo12}.

Anomalous transport has been studied extensively in laboratory plasmas, due to the importance of understanding and controlling the plasma losses, while it is comparatively less well known in geospace plasmas. In laboratory plasmas, transport was found to be nondiffusive already two decades ago, when it became clear that the random motion of particles in fluctuating electric and magnetic fields was analogous to motion in periodic and quasi-periodic potentials \citep{Zaslavsky89}, and that long ``ballistic'' displacements were possible in between the structure of nested magnetic surfaces which are characterised by broken surfaces called ``cantori'' \citep{Zaslavsky93}. The combination of temporal trapping and long displacements leads to anomalous diffusion which can be both subdiffusive and superdiffusive, depending on control parameters like the turbulence level and the particle energy \citep{Shlesinger93}. 

Many studies of anomalous transport in fluids and plasmas have been performed; for instance, \citet{Benkadda97} have shown that transport regimes including subdiffusion and superdiffusion can be obtained for passive tracer particles in a flow undergoing the transition to turbulence. \citet{Carreras01} have similarly explored the dynamics of tracer particles in turbulence models with avalanche transport. Numerical simulations of  fast ion transport in simple magnetized toroidal plasmas were performed by \citet{Gustafson12} and the fast ion transport perpendicular to magnetic field lines was quantified. Despite the simplicity of this system, the entire spectrum of suprathermal ion dynamics, from subdiffusion to superdiffusion, depending on beam energy and turbulence amplitude was observed. Interestingly, numerical simulation of particle transport in the solar wind turbulence find that perpendicular transport is either subdiffusive or normal, while parallel transport can be either superdiffusive or normal \citep{zimbardo06,Shalchi07}.

An overview of the theoretical approach to anomalous transport, including the use of non Gaussian statistics, the Hurst exponent, and continuos time random walks, plus the results of numerical simulations for laboratory and geospace plasmas is given by \citet{perrone12}. Here we discuss how to detect anomalous diffusion from \it in situ \rm measurements in space, with a focus on superdiffusion. 

In numerical simulations, one can follow a particle trajectory and numerically compute the mean square displacement; after that, a fit of $\langle \Delta x^2 \rangle$ versus time allows to distinguish between normal diffusion, subdiffusion, or superdiffusion. Another approach is based on the determination of the  Lagrangian velocity autocorrelation function, since ``thick'' power law tails are also indicative of superdiffusion \citep[\textit{e.g.},][]{mier08}. Again, this approach is feasible in numerical simulations but not in space measurements, where usually only the Eulerian correlation functions are available. Then, how to detect anomalous diffusion in space?

\begin{figure}[!htb] 
\vspace*{2mm}
\begin{center}
\includegraphics[width=0.60\textwidth]{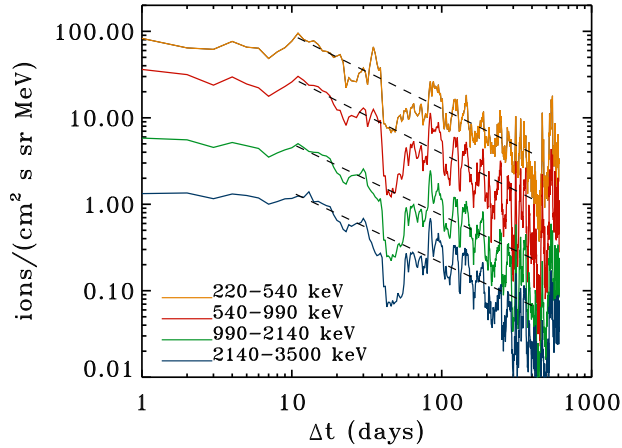}
\end{center}
\caption{Energetic ion fluxes measured by the LECP instrument onboard Voyager 2 upstream of the termination shock (energy as indicated). The dashed black lines represent the power law fit. Adapted from \citet{zimbardo12}. \label{figZim1}
}
\end{figure}

Already in 1974, by analysing the time profiles of nonrelativistic solar electron events, \citet{lin74} pointed out that nonrelativistic electrons exhibit a wide range of transport regimes, going from diffusive to ballistic. Indeed, an essential property of superdiffusive transport is that it is characterised by a non Gaussian statistics, both in the PDF of the diffusing particle (the \textit{propagator}) and in the probability of making a free path of a given length. Such a non Gaussian statistics influences the spatial distribution of superdiffusing particles, and hence the observed energetic particle time profile. In particular, superdiffusion corresponds to a Levy statistics, which implies that the propagator has the following power law shape, appropriate at some distance from the particle source at $x'$ \citep{zumofen93}, 
\begin{equation}
P(x-x',t-t') = {A_0\over (t-t')^{1/\alpha}} \left[{(t-t')^{1/\alpha}\over |x-x'|}  \ ,\right]^{\alpha + 1} \ ,
\label{eqzim11}
\end{equation} 
where $1<\alpha<2$, while a different expression holds for small values of $|x-x'|$ \citep{zumofen93}. Above, $A_0$ is a normalisation constant. As can be seen, the long distance propagator has a power law form, sharply different from the normal Gaussian propagator. The  density of energetic particles upstream of a shock can be obtained as the superposition of the particles accelerated at the shock during its propagation, and using the above propagator yields, for $t<0$ (\textit{i.e.}, upstream of the shock) \citep{perri07,perri08} 
\begin{equation}
f(E,t) \propto (-t)^{1-\alpha} \equiv (-t)^{-\nu} \ ,
\end{equation}
where $f(E,t)$ is the omnidirectional distribution function of particles of energy $E$. 
On the contrary, in the case of normal diffusion characterised by a Gaussian propagator, an exponential decay is obtained for $f(E,t)$ upstream of the shock. 
In other words, a power law time profile for energetic particles with slope $\nu=\alpha - 1 < 1$ is the signature of superdiffusion with anomalous diffusion exponent $\gamma=3-\alpha=2-\nu > 1$ \citep[][]{perri07,perri08,perrone12}.

\begin{figure}[!htb] 
\vspace*{2mm}
\begin{center}
\includegraphics[width=0.75\textwidth]{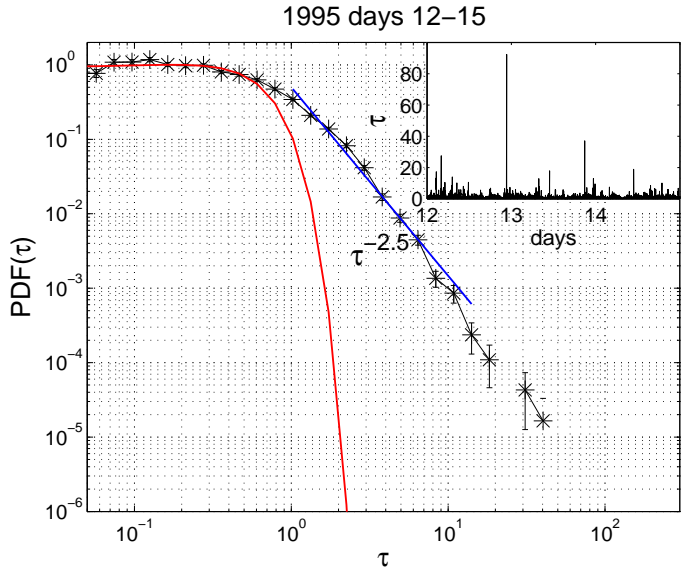}
\end{center}
\caption{PDF of scattering times measured by Ulysses in the fast solar wind in 1995 (asterisks). The blue line shows the power law fit, while the red line the Gaussian fit. The inset shows the time series of scattering times. Adapted from \citet{perri12b}. \label{figZim2}
}
\end{figure}

In order to extract information on the transport of energetic particles accelerated at the shock, one has to identify a ``clean'' shock crossing in the spacecraft data \citep{perri10a,perri10b}, then plot $f(E,\Delta t)$ upstream of the shock versus the time distance to the shock, $\Delta t = |t -t_{sh}|$.
A fit of $f(E,\Delta t)$ versus $\Delta t$ will reveal normal diffusion when the best fit is an exponential, and superdiffusion when the best fit is a power law. 
Using this technique as a diagnostic tool, \citet{perri07,perri08,perri09a} have shown that electron transport upstream of the shocks associated with corotating interaction regions (CIRs) detected by the Ulysses spacecraft in the solar wind at 4--5 AU is superdiffusive, with $\gamma \simeq 1.1$--$1.7$. On the other hand, analysing the Voyager 2 data of low energy particles, \citet{perri09b} have shown that ion transport upstream of the solar wind termination shock at 84 AU is superdiffusive, too, with $\gamma\simeq 1.3$, see Figure \ref{figZim1}. Let us stress again that a correct identification of a power law and its slope require great care, see Sect.~\ref{sec:spectra}.

We point out that the power law decay found in the energetic particle profiles upstream of CIR shocks is different from that envisaged by the numerical study by \citet{giacalone04}. In that study, the power law-like decay of $f(x,E,t)$ is due to the decrease of the magnetic fluctuations with increasing distance from the shock, so that pitch angle scattering decreases, too, and parallel transport becomes faster. However, \citet{perri12b} have shown that the  variance of the magnetic field components, $\sigma_i^2=\langle (B_i-\langle {B_i} \rangle )^2\rangle_{T}$ being $T$ the timescale for the average computation, measured by Ulysses at the electron resonance scale, which causes pitch angle scattering, do not vary with the distance from the shock. Therefore a constant level of magnetic fluctuations implies a constant transport regime, so that the energetic particle spatial profile, either an exponential decay or a power law with $\nu < 1$, can be used to discriminate between normal diffusion and superdiffusion. 

On the other hand, once superdiffusive transport is identified as explained above, one question arises: what it the physical origin of superdiffusion in the solar wind? A first answer can be given by considering that superdiffusion is due to a Levy statistics for the distribution $\psi(\ell)$ of free path lengths $\ell$. More precisely, the Levy random walk assumes 
$\psi(\ell) \sim |\ell|^{-1-\alpha}$ for large $|\ell|$; for constant velocity particles, this corresponds to a power law distribution of free path times, $\psi(\tau) \sim |\tau|^{-1-\alpha}$. For displacements along the magnetic field, one can relate the free path time to the pitch angle scattering time. In other words, a long pitch angle scattering time means that particles keep the same parallel velocity for a long time, thus performing the long displacements corresponding to a Levy random walk. The pitch angle scattering time is obtained from the normalised magnetic variance $ \sigma^2/B_0^2$ at the gyroresonant scale as $\tau=(\sigma^2/B_0^2)^{-1} \Omega^{-1}$ \citep{kennel66}, where $\sigma^2$ is the total magnetic variance near resonance and $\Omega$ is the particle gyrofrequency. \citet{perri12b} have shown that in the case of superdiffusive events in the solar wind, the distribution of pitch angle scattering times is a power law, see figure \ref{figZim2}, implying that long scattering times $\tau$ have a non negligible probability. Further, the power law slopes are between 2.4 and 3.8, values which are consistent, at least partly, with those required by the Levy random walk to have superdiffusion, \textit{i.e.}, $2<1+\alpha<3$. 

Therefore the analysis of the magnetic variances at the gyroresonant scale and the finding of a power law distribution of pitch angle scattering times with slope between 2 and 3 gives information on the microphysical processes leading to superdiffusion. On the other hand, this new data analysis technique calls for further investigations, since the scattering times determined by a spacecraft allow to compute the ``Eulerian distribution function'', while the random walk of the particles is determined by the scattering times seen along the particle trajectory, that is by the ``Lagrangian distribution function''. 

A different signature of anomalous transport, which is also intimately related to the concept of self-similarity, is the occurrence of long-range correlations. Models based on self-organized criticality \citep{newman96} and also various turbulence models predict the existence of such long-range correlations. By this, we mean a slow decay of the autocorrelation 
\begin{equation}
\lim_{\tau \rightarrow \infty} C(\tau) \sim \tau^{-\gamma} \ ,
\end{equation}
with $\gamma<1$. For such scalings, the integral time scale $T = \int_0^{+\infty} C(\tau) \; d\tau$ diverges and fundamental properties such as stationarity are lost. The main difficulty then consists in inferring such properties from finite and non-stationary records. Different approaches have been developed for that purpose, sometimes giving differing responses. One is the structure function approach, see Sect.~\ref{sec:strucf}. A different approach is rescaled range ($R/S$) analysis, in which the Hurst exponent quantifies the degree of persistence in a signal. We met this exponent before in the context of statistical self-similarity, see Eq.~(\ref{eq:selfsim}).

For a record of $N$ regularly sampled values with zero mean $\{y_1, y_2, \ldots, y_{N}\}$, we first create cumulative deviates $w_t = \sum_{i=1}^t y_i$ for $t=1, 2, \ldots, N$. From this we define the range as $R_t = \max(w_1, w_2, \ldots, w_t) - \min(w_1, w_2, \ldots, w_t)$. The rescaled range series then equals $(R/S)_t = R_t / \sigma_t$, where $\sigma_t$ is the standard deviation of the sequence $\{y_1, y_2, \ldots, y_t\}$. The following self-similar scaling holds
\begin{equation}
\lim_{\tau \rightarrow \infty} (R/S)_{\tau} \sim \tau^{-H} \ ,
\end{equation}
where $H$ is the Hurst exponent \citep{carreras98}. Persistence occurs when $0.5 < H < 1$, while $H=0.5$ indicates an uncorrelated process and $0 < H < 0.5$ indicates anti-persistence. For specific classes of processes, the value of $H$ is connected to that of the spectral index, and of the scaling exponent of the structure function \citep{gilmore02}.

There have been several applications of the rescaled range analysis to laboratory plasmas, see for example \citep[][]{yu03,tynan09}. The value of the Hurst exponent itself is not so informative; more interesting is the way it changes under different regimes, such as in the transition of from L to H-mode \citep{dudson05}. There have also been some applications to  geospace plasmas. \citet{kiyani07}, for example, used it to relate properties of the solar wind to the structure of the corona under different conditions of solar activity.

In many instances, $H$ has been found to be larger than 0.5, which is often considered as signature of self-organized criticality, or rather, driven self-organized criticality. However, as with all higher order statistical methods, there are also a number of pitfalls. In particular, $H$ must be determined within the appropriate range of time lags: $\tau$ must exceed local correlation time scales while being much smaller than the meso-scales of the system \citep{gilmore02}.

%%%%%%%%%%%%%%%%%%%%%%%%%%%%%%%%%%%%%%%%%%%%%%%%%%%%%%%%%%%%%%%%%%%%%%%%%%%%%%%%%%%%%%%

\section{Conclusions}
\label{sec:conclusions}

This brief overview of some advanced  methods for geospace and laboratory plasmas reveals the power of such methods, provided that they are used in connection with plasma physics theory, and that their limitations and pitfalls are properly understood.  One important, and yet missing, topic here is systems theory, which can give deep insight into the nonlinear dynamics of plasmas \citep{vassiliadis06}. One of the reasons for its relevance is the growing interest for system approaches, in which the coupling between different layers or regions becomes of comparable interest as the individual regions themselves.

Many time series analysis methods are relatively mature whereas spatio-temporal analysis methods often still are in their infancy. The CLUSTER multi-satellite mission is one of the rare examples wherein a coordinated effort toward the preparation of an experiment has led to new methodological developments. In most cases, however, innovation has occurred instead by strokes of serendipity or by knowledge transfer from nearby fields. This transfer has often been slower between the geospace and laboratory plasma communities than between the plasma and neutral fluid communities. Clearly, there remains a considerable issue in fostering interactions to the point where new ideas can spread more rapidly.

%%%%%%%%%%%%%%%%%%%%%%%%%%%%%%%%%%%%%%%%%%%%%%%%%%%%%%%%%%%%%%%%%%%%%%%%%%%%%%%%%%%%%%%

\bigskip

\subsection*{Acknowledgements}
{\small We all thank the International Space Science Institute (ISSI, Bern) for hospitality. In preparing this review we made extensive use of NASA's Astrophysics Data System.}

%%%%%%%%%%%%%%%%%%%%%%%%%%%%%%%%%%%%%%%%%%%%%%%%%%%%%%%%%%%%%%%%%%%%%%%%%%%%%%%%%%%%%%%

% \bibliographystyle{aps-nameyear}      
% \bibliography{methods}   

\end{document}